\documentclass[runningheads]{llncs}
\usepackage{balance}
\usepackage{algorithm}
\usepackage[noend]{algpseudocode}

\usepackage{amsmath}
\usepackage{amsfonts}
\usepackage{enumitem}
\usepackage{graphicx}
\usepackage{subcaption}
\usepackage{listings}
\usepackage{xcolor}
\usepackage{colortbl}
\usepackage{pgfplots}
\usepackage{changepage}
\usepackage{multirow}
\usepackage{url}
\usepackage{tablefootnote}
\usepackage{threeparttable}
\usepackage[mathscr]{euscript}
\usepackage[normalem]{ulem}
\usepackage{xspace}
\usepackage{hyperref} 
    \newcommand{\AICTHREE}{$\EuScript A\texttt{-IC3}$\xspace}
\usepackage[T1]{fontenc}
\usepackage{orcidlink}
\usepackage{marvosym}

\usepackage{graphicx}
\definecolor{mygood}{RGB}{102, 158, 65}
\definecolor{mybad}{RGB}{242, 98, 105}

\definecolor{myblue}{RGB}{66, 135, 245}

\makeatletter
\newcommand{\printfnsymbol}[1]{%
  \textsuperscript{\@fnsymbol{#1}}%
}
\makeatother

\begin{document}
\title{\AICTHREE: Learning-Guided \textbf{\uline{A}}daptive Inductive Generalization for Hardware Model Checking}
\titlerunning{\AICTHREE}
%

\author{Xiaofeng~Zhou\orcidlink{0000-0001-5878-3683}\thanks{Both authors contributed equally to this work.}\inst{1} \and
Guangyu~Hu\orcidlink{0000-0001-5077-8361}\printfnsymbol{1}\inst{1} \and
Hongce~Zhang\orcidlink{0000-0003-4001-264X}\inst{2} \and
Wei~Zhang\orcidlink{0000-0002-7622-6714}\textsuperscript{(\Letter)}\inst{1} }
\authorrunning{X. Zhou and G. Hu et al.}
\institute{The Hong Kong University of Science and Technology\\
\email{\{xzhoubu,ghuae\}@connect.ust.hk, wei.zhang@ust.hk} \and
The Hong Kong University of Science and Technology (Guangzhou)\\
\email{hongcezh@hkust-gz.edu.cn}
 }

\maketitle              
\begin{abstract}
%
The IC3 algorithm represents the state-of-the-art (SOTA)  hardware model checking technique,
owing to its robust performance and scalability. A significant body of research has focused on enhancing the solving efficiency of the IC3 algorithm, with particular attention to the inductive generalization process—a critical phase wherein the algorithm seeks to generalize a counterexample to inductiveness (CTI), which typically is a state leading to a bad state, into a broader set of states. This inductive generalization
is a primary source of clauses in IC3
and thus plays a pivotal role in determining the overall effectiveness of the algorithm.

Despite its importance, existing approaches often rely on fixed inductive generalization strategies, overlooking the dynamic and context-sensitive nature of the verification environment in which spurious counterexamples arise. This rigidity can limit the quality of generated clauses and, consequently, the performance of IC3.

To address this limitation, we propose a lightweight machine-learning-based framework that dynamically selects appropriate inductive generalization strategies in response to the evolving verification context. Specifically, we employ a multi-armed bandit (MAB) algorithm to adaptively choose inductive generalization strategies based on real-time feedback from the verification process. The agent is updated by evaluating the quality of generalization outcomes, thereby refining its strategy selection over time.

Empirical evaluation on a benchmark suite comprising 914 instances, primarily drawn from the latest HWMCC collection, demonstrates the efficacy of our approach. 
When implemented on the state-of-the-art model checker rIC3,
our method solves 26 to 50 more cases than the baselines and improves the PAR-2 score by 194.72 to 389.29.

\keywords{Hardware Formal Verification \and Model Checking \and Inductive Generalization \and Machine Learning.}
\end{abstract}
\section{Introduction}\label{sec:intro}

As the scale and complexity of modern \emph{RTL (register transfer level)} designs
continue to increase, simulation-based verification alone has become insufficient
for uncovering subtle design bugs.
Hardware model checking provides a formal and exhaustive means of verifying
safety properties by exploring all reachable states.
Among SAT-based techniques such as \emph{BMC (bounded model checking)}~\cite{BMC},
$k$-induction~\cite{KIND}, and IC3/PDR~\cite{IC3,PDR},
IC3 has emerged as a state-of-the-art algorithm due to its robustness
and its ability to incrementally construct inductive invariants
without explicit unrolling.

A key determinant of IC3's efficiency is the \emph{inductive generalization} step, in which a CTI \emph{counterexample to induction} is generalized into a clause
that blocks a larger set of unreachable states. Techniques such as CtgDown~\cite{CTG} and its extensions (e.g., stronger ICP/IC3 generalization~\cite{Winterer2021ICP} and EXCTG~\cite{EXCTG}) significantly improve generalization by performing more powerful reasoning about
\emph{CTGs (counterexamples to generalization)}. These works show that better generalization, in terms of the width and depth of blocked state space, can substantially accelerate IC3. However, they still depend on fixed or manually tuned strategies: the trade-off between generalization strength and computational overhead is decided once and for all, or adjusted only by handcrafted rules. For example, EXCTG further proposes DynAMic~\cite{EXCTG}, which dynamically switches between several predefined strategies based on the difficulty of blocking states, but the switching policy itself is static and hard-coded.

In practice, CTIs are generated under highly diverse circumstances: frame levels, clause densities, and the number of pending block tasks all evolve throughout the run of IC3.  Overly aggressive generalization may produce clauses that cannot be propagated, while overly conservative generalization may fail to block meaningful regions
of the state space, leading to slow convergence. The effectiveness of a generalization strategy is therefore highly context-dependent. Moreover, each generalization affects future CTIs, frame saturation, and clause propagation. From this perspective, inductive generalization is not an isolated subroutine,
but part of a sequential, non-stationary decision process that unfolds over the entire proof.

Recent learning-based verification approaches operate at different points in this design space. NeuroPDR~\cite{NeuroPDR} integrates neural networks into PDR to guide the proof search, and DeepIC3~\cite{DEEPIC3} uses a graph neural network to predict promising clauses for IC3. These methods demonstrate that machine learning can improve SAT/SMT-based model checking, but they rely on offline training, heavy models, and non-trivial inference cost. They are therefore not well-suited to be invoked millions of times inside IC3's inner inductive generalization loop. What is missing is a \emph{lightweight, online} mechanism that can adaptively select generalization strategies based on the evolving proof state, without requiring pre-collected datasets or expensive training.

\noindent\textbf{Our approach.}
In this work, we introduce \AICTHREE, a lightweight, learning-guided framework that adaptively selects inductive generalization strategies inside IC3. Rather than relying on a global fixed strategy or manually designed switching rules, \AICTHREE models generalization as a contextual multi-armed bandit problem. Before each generalization, the framework extracts a proof-aware context vector from IC3's internal state, including CTI characteristics (e.g., cube size, depth, activity) and global proof information (e.g., frame level, frame saturation, proof-obligation queue growth). A Proof-Aware LinUCB (PA-LinUCB) agent then selects among several predefined generalization strategies of varying aggressiveness. After generalization and intermediate clause propagation, the agent receives a reward that reflects the usefulness of the resulting clause, combining size reduction and push success, and updates its parameters online. Over time, \AICTHREE learns which strategies work best in which proof contexts, without any offline training.

\noindent\textbf{Why contextual bandits and LinUCB?}
Strategy selection in IC3 is inherently discrete, must incur negligible overhead, and must cope with non-stationary proof dynamics.
Contextual bandits provide exactly this combination of
(i) online learning without pre-collected training data,
(ii) low computational cost, and
(iii) the ability to condition decisions on rich context.
We adopt the LinUCB algorithm~\cite{LinUCB} as our decision agent: its linear value model matches the low-dimensional, structured features available in IC3, and its updates are simple enough to be integrated into industrial-scale model checking runs. The novelty of this work, therefore, does not lie in modifying LinUCB itself, but in identifying a proof-aware context and reward design, and an integration architecture that allows a contextual bandit to operate effectively inside
the recursive IC3 generalization loop.

This work makes the following contributions:
\begin{itemize}
    \item We formulate inductive generalization in IC3 as a
    contextual decision-making problem and propose \AICTHREE\footnote{Implementation available on Zenodo \url{https://zenodo.org/records/19811923}}, the first learning-based adaptive controller for generalization strategy selection in IC3.
    \item We design a proof-aware context representation and a push-aware reward function that jointly capture the local difficulty of generalizing a CTI and the global progress of the proof, enabling a standard LinUCB agent to provide meaningful guidance with negligible overhead.
    \item We integrate our method into rIC3~\cite{ric3}, a state-of-the-art IC3
implementation, and evaluate it on 914 HWMCC benchmark instances. The results
show consistent improvements in solved instances and PAR-2 scores over multiple
baselines, demonstrating the practical benefit of learning-guided,
context-aware generalization.
\end{itemize}

By elevating inductive generalization from a static heuristic to an
adaptive, learning-guided process, \AICTHREE complements existing
generalization techniques such as CTG/EXCTG and opens a new direction
for integrating lightweight machine learning techniques into
symbolic model checking. 

The remainder of this paper is organized as follows.
Section~\ref{sec:pre} introduces the background of IC3 and the
multi-armed bandit method.
Section~\ref{sec:method} presents the proposed \AICTHREE framework and the
proof-aware contextual bandit formulation.
Section~\ref{sec:res} reports our experimental results on rIC3 over the HWMCC'20, HWMCC'24 and HWMCC'25 benchmark suite.
Section~\ref{sec:related_works} discusses related work, and
Section~\ref{sec:conclusion} concludes the paper.

\section{Preliminaries}\label{sec:pre}

\subsection{Transition Systems and IC3/PDR}

We model a hardware system as a Boolean transition system
$\langle V, T(V,V'), I(V)\rangle$, where $V$ is a finite set of state
variables, $V'$ denotes the corresponding next-state variables,
$I(V)$ characterizes the initial states, and $T(V,V')$ is the
transition relation.
A \emph{literal} $l$ is a variable $v \in V$ or its
negation $\neg v$.
A \emph{clause} $c$ is a disjunction of literals, and
its negation is a \emph{cube} $s$, representing a conjunction of
literals.
We write $P(V)$ for the safety property to be verified; states
violating $P$ are called \emph{unsafe}.

\begin{algorithm}[htbp]
\caption{The IC3 Algorithm (A simplified version with details omitted)}
\label{alg:ic3}
\begin{algorithmic}[1]
\State \textbf{Input:} Transition function $\mathcal{T}$, initial states $\mathcal{I}$, property $P$
\State \textbf{Output:} The proving result of $P$

\State Initialize frames $F_i$ for $i \geq 0$ such that $F_0 = \mathcal{I}$, $k \gets 1$
\State \textcolor{blue}{$a \gets \text{GENERALIZATION\_STRATEGY} $} \label{ic3:line:glb_str}

\While{True} \label{line:ic3:outer_loop_start}
    \While {SAT($F_k \land \neg P$)} \label{line:ic3:strengthen_loop_start}
        \State $s \gets \text{GetModel()}$
        \If {$\neg$ Block($s$, $k$, $a$)} \label{line:ic3:block}
            \State \Return UNSAFE
        \EndIf
    \EndWhile \label{line:ic3:strengthen_loop_end}

    \For {$1 \leq i < k$} \label{line:ic3:propagate_loop_start}
        \If{Propagate($F_i$, $F_{i+1}$) = CONVERGED}  \label{line:ic3:propagate_check}
            \State \Return SAFE
        \EndIf
    \EndFor \label{line:ic3:propagate_loop_end}
    \State $k \gets k + 1$ \label{line:ic3:outer_loop_end}
\EndWhile

\Procedure{Block}{$s$, $k$, $a$}
    \If{$k = 0$}
        \State \Return False \Comment{Find a trace to unsafe} \label{line:ic3_block:unsafe}
    \EndIf
    \While{SAT($F_k \land \neg s \land \mathcal{T} \land s'$)} \label{line:ic3_block:inductive} 
        \State $s_p \gets $GetModel()
        \If{$\neg$ Block($s_p$, $k-1$, $a$)} \label{line:ic3:block_recursive}
            \State \Return False \Comment{Failed to block on frame $k$}
        \EndIf
    \EndWhile
    \State \textcolor{blue}{$c \leftarrow \text{Generalize}(s, a)$}
    \State $F_i \leftarrow F_i \wedge c$ for $1 \leq i \leq k$
    \State \Return True
\EndProcedure

\end{algorithmic}
\end{algorithm}

IC3/PDR~\cite{IC3,PDR} decides whether unsafe states are reachable
by constructing an inductive invariant $\mathit{INV}(V)$ that satisfies:
\begin{equation}
\label{eq:inv}
\begin{aligned}
  &I(V) \rightarrow \mathit{INV}(V),\\
  &\mathit{INV}(V) \wedge T(V,V') \rightarrow \mathit{INV}(V'),\\
  &\mathit{INV}(V) \rightarrow P(V).
\end{aligned}
\end{equation}
If such an $\mathit{INV}$ exists, it over-approximates the set of
reachable states while excluding all unsafe states, thereby proving
safety.

IC3 maintains a sequence of frames $F_0, F_1, \ldots, F_N$, where each
frame $F_i$ is represented as a conjunction of clauses:
\[
  F_i = c_1 \wedge c_2 \wedge \cdots \wedge c_{n_i}.
\]
$F_0$ is initialized to $I(V)$, and the frames satisfy:
\begin{equation}
\label{eq:frames}
\begin{aligned}
  &F_0(V) \rightarrow I(V), \\
  &F_i(V) \rightarrow F_{i+1}(V) \quad \text{for all } i < N, \\
  &F_i(V) \wedge T(V,V') \rightarrow F_{i+1}(V') \quad \text{for all } i < N, \\
  &F_i(V) \rightarrow P(V) \quad \text{for all } i.
\end{aligned}
\end{equation}
Intuitively, $F_i$ over-approximates the set of states reachable from
$I$ in at most $i$ steps, and the sequence becomes progressively
less restrictive, i.e., $F_0 \Rightarrow F_1 \Rightarrow \cdots \Rightarrow F_N$.
When two adjacent frames coincide, $F_k \equiv F_{k+1}$, the conjunction
$F_k$ is an inductive invariant that satisfies~\eqref{eq:inv}, and the property is proved.

Algorithm~\ref{alg:ic3} presents the pseudocode of the IC3 procedure. At a high level, IC3 iteratively searches for counterexamples and refines the frames by learning clauses. \textbf{Block}: given a candidate bad predecessor, IC3 derives a generalized clause to exclude it and adds the clause to the corresponding frame(s). \textbf{Propagate}: IC3 then pushes learned clauses to higher frames whenever possible, which accelerates convergence.
If a SAT query discovers a state $s$ that reaches an unsafe state
within $i$ steps while $s \models F_i$, $s$ is a CTI at frame $F_i$.
IC3 then tries to \emph{block} $s$ by finding a clause $c$ such that
$c$ excludes $s$, is inductive relative to $F_{i-1}$, and can be pushed
forward to higher frames.
We call a pair \((s,i)\), where cube \(s\) must be blocked at frame \(F_i\), a proof obligation (PO). 
The PO queue stores pending blocking tasks generated during recursive blocking. 
This blocking step is performed recursively on predecessor states and generates POs that must be blocked at lower frames.
Figure~\ref{fig:context_vec} illustrates the IC3 frame structure together with
the local (CTI-related) and global (frame-related) quantities introduced above.
These quantities will later serve as the building blocks of our proof-aware
context vector in Section~\ref{sec:method}.

\subsection{Inductive Generalization and CTG-Based Methods}

The clause used to block a CTI is typically obtained by
\emph{generalizing} the CTI cube $s$ with respect to the current frame.
Generalization seeks to drop literals from $s$ while preserving
inductiveness relative to $F_{i-1}$.
The resulting clause determines how many states are blocked and
how easily it can be propagated to higher frames.
Two failure modes are well known in the IC3 literature~\cite{CTG}:
\begin{itemize}
  \item \textbf{Over-generalization}: dropping too many literals yields a
  strong clause that is not valid at higher frames and therefore cannot
  be pushed, inflating the queue of proof obligations that must be
  blocked.
  \item \textbf{Under-generalization}: keeping too many literals yields a
  very weak clause that blocks only a tiny neighbourhood of the CTI,
  resulting in many similar CTIs and slow convergence.
\end{itemize}

CtgDown~\cite{CTG} improves generalization by introducing
CTGs: when a literal dropping attempt fails, the model returned by the SAT solver is
treated as a CTG whose predecessors are explored to
search for more general clauses.
EXCTG~\cite{EXCTG} further strengthens this idea by exploring CTGs recursively and carefully tuning the recursion depth, the number of CTGs considered, and the number of literal-dropping attempts.
These methods demonstrate that better generalization---in terms
of both width (states blocked) and depth (frames reached)---can
substantially accelerate IC3, but they still rely on fixed or
manually tuned settings for these parameters.

In this work, we build upon CTG-based generalization and treat
the configuration of these parameters (e.g., recursion depth, CTG
budget, literal-dropping effort) as a strategy whose strength
varies from conservative to aggressive.
Our goal is to adaptively select such strategies according to the
current proof context, rather than use a single fixed configuration
throughout the run.

\subsection{Multi-Armed Bandits and LinUCB}

The \emph{multi-armed bandit} (MAB) problem models an agent
that repeatedly chooses an arm $a$ from a finite set $\mathcal{A}$ and
receives a stochastic reward $r$.
The objective is to maximize the cumulative reward (or equivalently,
to minimize regret) by balancing exploration of different arms and
exploitation of arms known to perform well.
In contextual bandits, the agent additionally observes a
feature vector $x_t$ at each round $t$, representing the current
context, and selects an arm based on both the context and
past observations.

Among many contextual bandit algorithms, \emph{LinUCB (Linear Upper Confidence Bound)}~\cite{LinUCB} assumes that the expected reward of each arm is a linear function of the context. For each arm $a \in \mathcal{A}$, LinUCB maintains a parameter vector and a (regularized) covariance matrix, and selects arms using
an upper-confidence-bound rule that trades off estimated reward and uncertainty.
Its per-round update cost is low and depends only on the dimension
of the context vector, making it attractive for settings such as IC3, where decisions are made extremely frequently, and overhead must be tightly controlled.

In \AICTHREE, we instantiate the contextual bandit with LinUCB
and view each generalization configuration as an arm.
The formal bandit formulation, the design of the proof-aware context
vector, and the LinUCB integration into IC3’s blocking loop are given in Section~\ref{sec:method}.

\section{Method}\label{sec:method}


\subsection{A-IC3 Overview}\label{sec:method:overview}

\begin{figure}[htb]
    \centering
    \includegraphics[width=0.9\linewidth,alt={Overview of A-IC3. The left panel illustrates a multi-armed bandit selecting among four generalization arms over four steps, with selected arms highlighted and a gradient indicating increasing generalization strength. The right panel shows the decision pipeline: proof-context features are extracted from the hardware model checker, passed to a PA-LinUCB agent, used to select the best arm for clause generalization, and then updated using feedback from the solving progress.}]{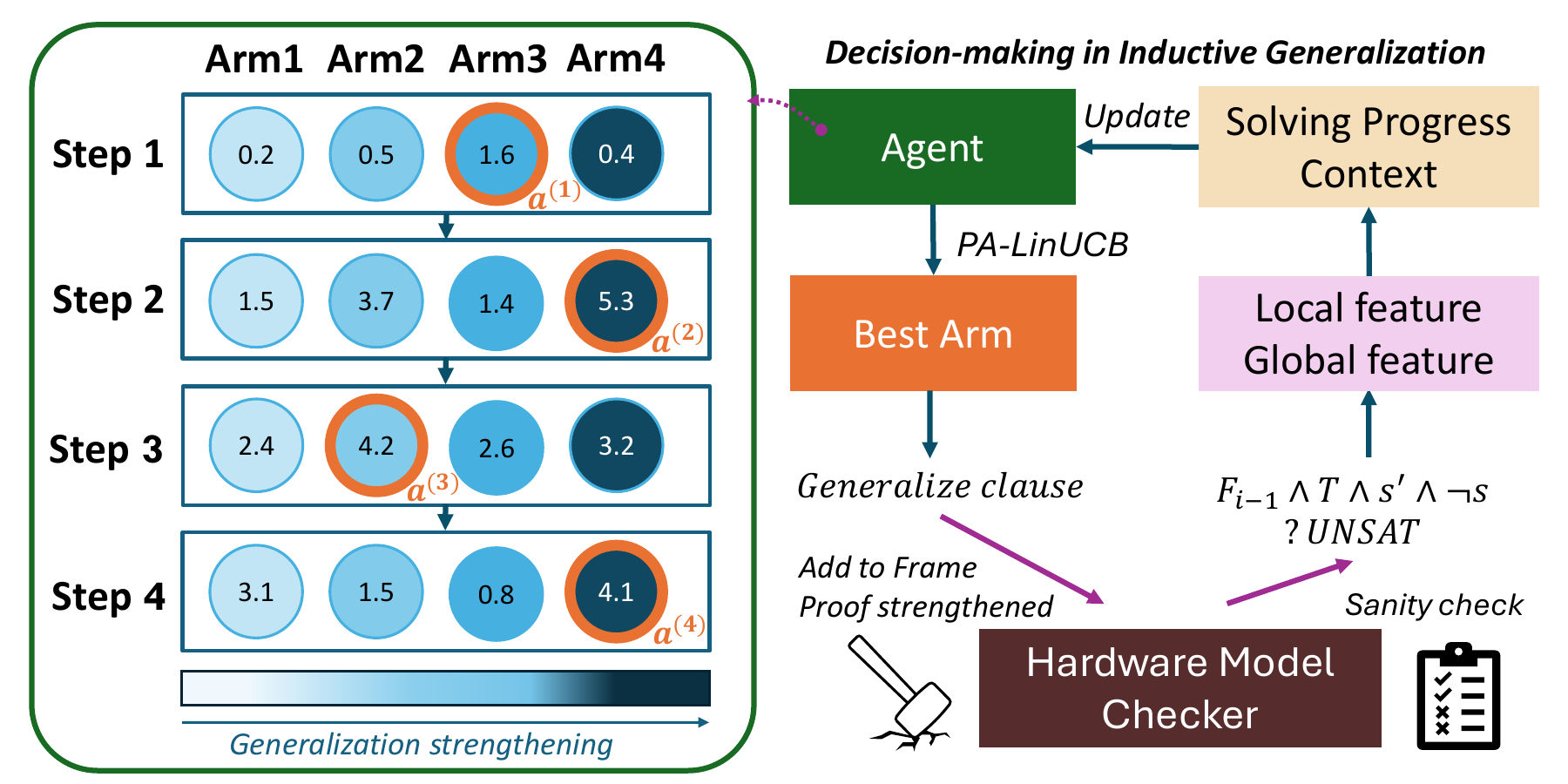}
    \caption{Overview of \AICTHREE.}
    \label{fig:overview}
\end{figure}

Current IC3 implementations face a fundamental limitation in generalization strategy selection: fixed strategies cannot adapt to the highly dynamic nature of proof contexts. This rigidity leads to systematic inefficiencies—over-generalization wastes computational resources on hard-to-push clauses, while under-generalization fails to discover sufficiently broad blocking clauses, severely hindering proof convergence.

\AICTHREE overcomes this limitation by treating generalization strategy selection as a contextual multi-armed bandit problem, as illustrated in Figure~\ref{fig:overview}. The left side of the figure demonstrates the multi-armed bandit framework in action: at each step, our PA-LinUCB agent, which is inspired by LinUCB~\cite{LinUCB} and tailored for the environment of the IC3 algorithm observes the current proof context and selects among different generalization strategies (represented as arms with varying reward values). The agent continuously learns from the outcomes—for instance, in Step 1, Arm3 with the best score 1.6 is selected and marked with $a^{(1)}$, the corresponding strategy will be used in the next generalization in the IC3-based hardware model checker. In the next step, similarly, after the agent is updated, we observe that Arm4 achieves the highest score of 5.3, indicating that it will be better to use a more aggressive generalization. The Arm4 is then selected and marked with $a^{(2)}$. The agent will progressively refine selections that select different arms in subsequent steps according to the solving progress.

The right side of Figure~\ref{fig:overview} shows our decision-making pipeline for inductive generalization. The agent leverages both local and global features from the solving progress context to make informed strategy selections. Each generalization strategy is encoded as a bandit arm, collectively orchestrated by our PA-LinUCB agent that adapts its choices based on real-time verification context inside the hardware model checker.

Following each generalization-and-propagation iteration, the agent evaluates the quality of the generated clause through internal metrics of IC3
and computes a reward based on this evaluation. This feedback enables adaptive learning over time, allowing the agent to refine its strategy selection policy in response to evolving proof dynamics, as demonstrated by the iterative improvement in arm selection across the four steps shown in Figure~\ref{fig:overview}.

\subsection{Contextual MAB Formulation}

As stated earlier, we formulate the generalization strategy selection problem as a contextual MAB problem with the context from IC3. In order for the agent to make informed selections, we need to convey the real-time statistical proving information to the agent. This information is encoded as a context vector.  In this setting, at the $t$-th call of the generalization procedure, we provide the agent with a context vector $\mathbf{x}_t \in \mathbb{R}^d$ that encodes the current state of the IC3 proof process, selects an arm (strategy) $a_t$ from a finite set of arms $\{a_1, a_2, \ldots, a_K\}$, and receives a reward $r_t$ based on the effectiveness of the chosen strategy on generalization.

\begin{figure}
    \centering
    \includegraphics[width=0.9\linewidth,
    alt={Context-vector extraction from an IC3 proof state. The diagram shows nested frames $F_0$ to $F_4$, a CTI at frame $F_2$, predecessor and successor CTIs connected through a proof-obligation queue, and an unsafe-property target. The extracted local features are CTI depth 3, cube size 4, activity score 0.85, and proof-obligation queue length 2; the extracted global features are frame level 2, frontier frame level 4, and frame saturation 4.}
]{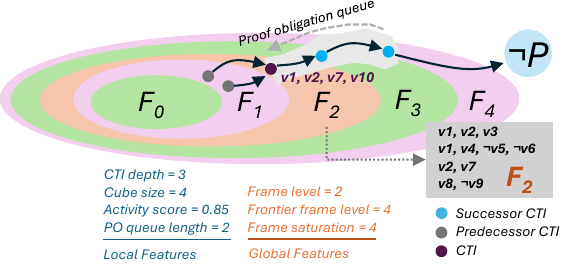}
    \caption{Context vector extraction from IC3 proof state. The figure illustrates IC3's frame structure where each frame $F_i$ contains clauses (e.g., ``$v1, v2, v3$'' represents one clause) that over-approximate reachable states at step $i$. The purple circle represents a CTI at frame $F_2$ with cube content $v_1, v_2, v_7, v_{10}$. The arrows show predecessor-successor relationships. The context vector combines local features (CTI depth, cube size, activity score) with global features (frame levels, frame saturation, proof obligation queue length) to guide MAB strategy selection.}
    \label{fig:context_vec}

\end{figure}

\textbf{Context Vector Design.} The context vector $\mathbf{x}_t \in \mathbb{R}^d$ encodes the current proof state through carefully selected features. Figure~\ref{fig:context_vec} shows how \AICTHREE extracts context before the generalization procedure. This context design balances \emph{local proof characteristics} with \emph{global progression indicators} (indicated by ``local features'' and ``global features'' in Figure~\ref{fig:context_vec}), incorporating both absolute metrics (for global positioning) and relative metrics (for adaptive scaling). 

The \emph{local context component} captures information that is directly tied to the current CTI and its immediate blocking workload, including the \textbf{depth}, \textbf{cube size}, \textbf{activity score}, and the \textbf{PO-queue length} when the CTI is encountered.
The activity score, defined in~\cite{EXCTG}, is a normalized measure of how frequently the CTI under generalization is encountered during proof search; we denote it as $\hat{x}_{a}$.
The cube size is the number of literals in the cube to be generalized.
The depth is the recursion depth of the \textsc{Block} call stack (Algorithm~\ref{alg:ic3}) at which the CTI is discovered.
The PO-queue length indicates how many proof obligations are pending alongside the current CTI, reflecting the \emph{local blocking pressure} at the moment of generalization.
Together, these features characterize the difficulty of handling the current CTI.
For example, a shallow CTI with a high activity score often suggests a complex and still-uncovered predecessor region, where a strategy with more generalization effort may be beneficial.
In Figure~\ref{fig:context_vec}, when blocking the purple CTI on frame $F_2$ (cube size $=4$), the proof-obligation queue already contains two successor obligations from higher frames $F_3$ and $F_4$, indicating a PO-queue length of $2$.
Moreover, the current CTI is three transitions away from the bad states $\neg P$, so the \textsc{Block} call-stack depth is $3$.
For normalization, we scale the depth by the relative frame index with respect to the frame where the PO queue is established, denoted as $\hat{x}_{d}$, normalize the cube size by the running average cube size observed so far, denoted as $\hat{x}_{c}$, and normalize the PO-queue length by its maximum observed value, denoted as $\hat{x}_{q}$.

The \emph{global context component} summarizes the overall proof progress, including the current frame level, the frontier frame level, the \textbf{frame saturation}, and the \textbf{growth rate} of the PO queue. 
Frame saturation is quantified as the number of clauses in the current frame, followed by normalization for learning. 
The PO-queue growth rate is defined as the normalized difference between the current PO-queue length and that recorded at the previous generalization call.
In Figure~\ref{fig:context_vec}, we generalize a CTI at frame $F_2$, the frontier is $F_4$, and frame $F_2$ contains $4$ clauses; thus the corresponding global features are: current frame level $=2$, frontier level $=4$, and (raw) frame saturation $=4$.
The growth rate of the PO queue reflects the recent global trend of the search: whether proof obligations are being discharged effectively (queue shrinking) or accumulating (queue growing).
For normalization, we represent the current frame level as a ratio to the frontier, denoted as $\hat{x}_{cf}$, normalize frame saturation by dividing it by $100$, denoted as $\hat{x}_{s}$, and normalize the PO-queue growth rate by its maximum observed value, denoted as $\hat{x}_{qr}$.

In summary, we define the context vector $\mathbf{x}_t \in \mathbb{R}^{8}$ of \AICTHREE at time $t$ as $\mathbf{x}_t = [\hat{x}_{d}, \hat{x}_{c}, \hat{x}_{q}, \hat{x}_{cf}, \hat{x}_{s}, \hat{x}_{qr},\hat{x}_a, b]$, we set $b = 1.0$.



Such a multi-faceted representation enables the MAB agent to distinguish between scenarios favoring aggressive versus conservative generalization. 
The level-related information can inform the agent about the rough progress of the proof, while frame saturation indicates whether there are too many clauses in the current frame, which may slow down the blocking and generalization procedures.
\textcolor{black}{A long proof obligation queue or a rapid increase in its length may suggest adopting a more aggressive generalization strategy, which can help reduce the generation of new CTIs and CTGs.} 
A higher saturation level means that the frame is highly restricted, possibly requiring the agent to pick up a strategy with less effort to prevent computational effort. It should be noted that, to better fuse the IC3 context information into the MAB agent, we perform normalization before loading the components into the context vector, and a bias term $b$ for flexibility is attached at the end. 


\textbf{Strategic Arm Configuration.}
Our framework employs a spectrum of generalization strategies that differ in the aggressiveness of searching for clauses, including \textbf{Basic}, \textbf{Conservative}, \textbf{Balanced}, and \textbf{Aggressive}. These strategies are built upon the CtgDown generalization approach~\cite{CTG}.
Each strategy controls the intensity of the 
generalization process through different configurations of CTG attempt counts, recursion depths, and literal-dropping constraints.

The four strategic arms employ progressively more aggressive parameter configurations:

\begin{itemize}
    \item \textbf{Basic} is selected for low-activity proof obligations at constrained frames where minimal generalization is preferred. It disables CTG exploration entirely (setting attempt count and recursion depth to 0) while allowing extensive literal dropping trials to avoid wasting computation on simple cases.
    \item \textbf{Conservative} applies cautious exploration suitable for early proof stages. It uses limited CTG attempts (1-3 attempts) with shallow recursion depths (1-2 levels) while permitting substantial literal dropping trials to balance exploration with efficiency.
    \item \textbf{Balanced} represents standard generalization for typical verification scenarios. It employs moderate CTG attempts (1-5 attempts) with expanded recursion depths (1-4 levels) and generous literal dropping allowances for comprehensive but controlled exploration.\looseness=-1
    \item \textbf{Aggressive} enables extensive exploration for high-activity proof obligations at advanced frames, utilizing substantial CTG attempts (2-4 attempts) with deep recursion levels (up to 8 levels), while maintaining flexible literal dropping policies to maximize generalization potential.
\end{itemize}

These tiered configurations enable adaptive generalization across diverse verification contexts, allowing the framework to dynamically find the strategy of appropriate exploration depth and width based on the proof obligation and the global environment of IC3 proving. Note that the ranges above describe the overall strategy design space; in \AICTHREE, we discretize this space into a finite set of arms by instantiating each conservative/balanced/aggressive family with concrete settings.

\subsection{PA-LinUCB Algorithm for Strategy Selection}

\begin{algorithm}[h]
\caption{MAB-based IC3 generalization}
\label{alg:mab_ic3}
\begin{algorithmic}[1]
\State \textbf{Global:} MAB agent $\mathcal{A}$, Frames $\{F_i\}$, Time stamp $t$.
\State \textbf{Initialize:} $\mathcal{A}.A_{a_i} =\mathbf{I}, \mathcal{A}.b_{a_i} = \mathbf{0}$ for $1 \leq i \leq K$, $t = 1$.
\Procedure{Block}{$s$, $k$}
    \If{$k = 0$}
        \State \Return False \Comment{Find a trace to unsafe} \label{line:mab_block:unsafe}
    \EndIf
    \While{SAT($F_k \land \neg s \land \mathcal{T} \land s'$)} \label{line:mab_block:inductive} 
        \State $s_p \gets $GetModel()
        \If{$\neg$ Block($s_p$, $k-1$)} \label{line:mab:block_recursive}
            \State \Return False \Comment{Failed to block on frame $k$}
        \EndIf
    \EndWhile
    \State \textcolor{blue}{$\mathbf{x}_t \leftarrow \text{ExtractContext}(s, \{F_i\})$} \Comment{Context extraction} \label{line:mab_block:mab_start} 
    \State \textcolor{blue}{$a_t \leftarrow \mathcal{A}.\text{SelectArm}(\mathbf{x}_t)$} \Comment{PA-LinUCB selection} \label{line:mab_block:select_arm} 
    \State \textcolor{blue}{$c \leftarrow \text{Generalize}(s, a_t)$} \label{line:mab_block:generalize_mab} 
    \State $\text{pushed\_frame} \leftarrow \text{PushClause}(c, k)$ \Comment{Intermediate push} \label{line:mab_block:inter_push}
    \State \textcolor{blue}{$r_t \leftarrow \text{ComputeReward}(c, \text{pushed\_frame}, \{F_i\})$} \label{line:mab_block:reward}
    \State \textcolor{blue}{$\mathcal{A}.\text{UpdateArm}(a_t, \mathbf{x}_t, r_t)$} \Comment{LinUCB update} \label{line:mab_block:mab_end}
    \State \Return True
\EndProcedure

\end{algorithmic}
\end{algorithm}

As discussed in Section~\ref{sec:method:overview}, we develop PA-LinUCB (Proof-Aware LinUCB) as our decision agent, specifically tailored for IC3's generalization context. PA-LinUCB deeply integrates with IC3's proof structure, leveraging proof-specific contextual information to guide generalization strategy selection. 
This integration provides more information to the agent about the dynamic proof environment and adapt its decisions based on the current proof state, frame structure, and clause characteristics.

PA-LinUCB maintains a linear model for each generalization strategy, assuming that the expected reward can be expressed as a linear combination of proof-aware features. This assumption aligns well with IC3's structured proof process, where generalization effectiveness often correlates linearly with specific proof characteristics such as clause size, frame depth, and propagation potential according to our observations.

At each generalization step, PA-LinUCB computes the upper confidence bound for each strategy $a$:

\begin{equation}
\text{PA-LinUCB}_a(\mathbf{x}_t) = \boldsymbol{\theta}_a^T \mathbf{x}_t + \alpha \sqrt{\mathbf{x}_t^T \mathbf{A}_a^{-1} \mathbf{x}_t}
\label{eq:pa_linucb}
\end{equation}

where $\mathbf{x}_t$ represents the proof-aware context vector extracted from the current IC3 state, $\boldsymbol{\theta}_a$ captures the strategy's learned parameters, $\mathbf{A}_a$ denotes the regularized covariance matrix associated with strategy $a$, and $\alpha$ controls the exploration level. The strategy with the highest score is selected as:
\begin{equation}
a_t = \arg\max_{a \in \mathcal{A}} \text{PA-LinUCB}_a(\mathbf{x}_t)
\end{equation}

After observing the generalization result, PA-LinUCB updates the selected strategy's parameters to incorporate the proof-specific feedback:

\begin{align}
\mathbf{A}_{a_t} &\leftarrow \mathbf{A}_{a_t} + \mathbf{x}_t \mathbf{x}_t^T \\
\mathbf{b}_{a_t} &\leftarrow \mathbf{b}_{a_t} + r_t \mathbf{x}_t \\
\boldsymbol{\theta}_{a_t} &\leftarrow \mathbf{A}_{a_t}^{-1} \mathbf{b}_{a_t}
\end{align}

The reward $r_t$ captures both immediate generalization effects and long-term proof progress, enabling PA-LinUCB to learn from IC3's complex feedback signals. The design of the reward $r_t$ will be discussed in detail in Section~\ref{sec:reward}. $b_{a_t}$ is the weighted reward vector after performing the action $a_t$, and is derived via scaling $r_t$ by their corresponding context vectors.

Algorithm~\ref{alg:mab_ic3} presents the deep integration of PA-LinUCB with IC3's blocking procedure. The key innovation is around lines~\ref{line:mab_block:mab_start}-\ref{line:mab_block:mab_end}, where proof-aware context extraction (Line~\ref{line:mab_block:mab_start}), strategy selection (Line~\ref{line:mab_block:select_arm}), and parameter updates (Line~\ref{line:mab_block:mab_end}) are seamlessly embedded within IC3's core block and generalization process. An intermediate push is conducted via the function PushClause, pushing the generalized clause into higher frames during the block of the current frame, in order to derive the effect of the generalized clause (Line~\ref{line:mab_block:inter_push}). This tight integration ensures that PA-LinUCB operates with full awareness of IC3's proof state and can adapt its strategy selection based on the evolving proof context.

\subsection{Reward Function Design}\label{sec:reward}

The reward function serves as the critical feedback mechanism that guides our MAB agent toward effective generalization strategies. As introduced in earlier sections, fixed-strategy generalization often results in over-generalization and under-generalization. Our design explicitly addresses this generalization strength balancing problem in the reward calculation part. 
\AICTHREE uses a multi-component reward structure. We decompose the reward into the following complementary components that capture different aspects of generalization quality.

\begin{itemize}[noitemsep, topsep=0pt, partopsep=0pt]
    \item \textbf{Size reduction quality} $R_s$ directly reflects the clause simplification effectiveness, indicating the ability of state refinement of the generalized clause. It is calculated as the proportion of clause size reduction relative to the original size. If the size increases after generalization, we pose a penalty factor $\beta > 1$.
    \item \textbf{Push quality} $R_p$ evaluates the propagation success of clause. It is calculated in two conditions: if the clause is pushed to higher frames successfully, $R_p$ equals the ratio of the push distance and the maximum possible push distance (the distance between the frame of proof obligation and the frontier frame); If the clause cannot be pushed to another frame at all, we set a constant penalty to $R_p = p_p$.
    \item \textbf{Bonus} $R_b$ highlights extreme outcomes of a generalization step in a reward signal (e.g., a clause reaches the frontier or exhibits over-generalization). The bonus helps the agent react quickly by discouraging consistently unproductive arms and reinforcing promising ones. Each event has a polarity $s_i\in\{+1,-1\}$ (bonus/penalty) and an importance level $\ell(i)\in\{\mathrm{H,M,L}\}$. We use qualitative magnitudes satisfying $\gamma_{\mathrm{H}}>\gamma_{\mathrm{M}}>\gamma_{\mathrm{L}}>0$ and define
    \begin{equation}
    R_b \;=\; \sum_i s_i\,\gamma_{\ell(i)}\,\mathbb{I}[E_i],
    \label{eq:Rb}
    \end{equation}
    
    where $\mathbb{I}[E_i]\in\{0,1\}$ indicates whether event $E_i$ is triggered. The events are summarized in Table~\ref{tab:bonus}. When multiple events are triggered, their contributions are accumulated.
    
    We set $E_{\text{front}}$ to High because pushing a clause to the current frontier is a decisive success signal: such a clause does not hinder subsequent clause propagation toward the proof frontier, and it can effectively constrain the state space by strengthening the reachable frames. We set $E_{\text{size1}}$ to Medium since a size-$1$ clause is often very strong, yet aggressive size reduction may occasionally introduce over-generalization and lead to adverse effects, so we avoid over-rewarding this signal. We treat $E_{\text{high}}$ as a weaker variant of $E_{\text{front}}$: pushing to a relatively high frame indicates progress but is less conclusive and occurs more frequently than reaching the frontier, hence it receives a \textbf{Low} importance. We set $E_{\text{ideal}}$ to Medium because it captures a well-balanced outcome: the generalization produces a clause that is both strong and sufficiently pushable, even if it is not the most extreme along either dimension. Empirically, repeatedly triggering $E_{\text{ideal}}$ indicates that the selected arm constitutes a consistently feasible and effective strategy under the current proving context, and thus deserves a moderate importance weight. $E_{\text{over}}$ captures over-generalization: although the derived clause is strong, it is barely pushable to higher frames. This outcome suggests that the current strategy is mismatched to the present proving context.

\end{itemize}

\begin{table}[t]
\centering
\caption{Events used in the bonus term $R_b$.}
\label{tab:bonus}
\begin{tabular}{|c|p{0.66\linewidth}|c|c|}
\hline
\textbf{Tag} & \textbf{Event} & $s_i$ & $\ell(i)$ \\
\hline
$E_{\mathrm{front}}$ &
Frontier push: Clause is pushed to frame $F_k$. & $+1$ & High \\
$E_{\mathrm{size1}}$ &
Complete generalization: clause size is $1$ (one literal). & $+1$ & Medium \\
$E_{\mathrm{high}}$ &
High-level push: the clause is pushed to a frame $k' > 0.7k$
(where $k$ is the frontier frame). & $+1$ & Low \\
$E_{\mathrm{ideal}}$ &
Ideal generalization: $R_s > 0.5$ and $R_p > 0.3$. & $+1$ & Medium \\
$E_{\mathrm{over}}$ &
Over-generalization: $R_s > 0.7$ and $R_p < 0.1$. & $-1$ & Medium \\
\hline
\end{tabular}
\end{table}

Finally, we can derive the overall reward, where $w_s$ and $w_p$ are weights for balancing the size reduction and push quality:

\begin{equation}
    r_t = w_s R_{s} + w_pR_{p} + R_{b}.
\label{eq:reward}
\end{equation}

\section{Evaluation}\label{sec:res}

\subsection{Experimental Setup}

All experiments were conducted on a high-performance computing cluster consisting of twenty nodes. Each node is equipped with 64 CPU cores, specifically Intel Xeon Gold 6338 CPUs operating at 2.00GHz. The system runs Red Hat Enterprise Linux Server version 7.9 and provides 1007GB of memory per node. Each invocation of the model checker was executed with one dedicated CPU and memory resources allocated by the scheduler to ensure consistency and reproducibility. The model checker is executed in a single thread, and no portfolio is included. The weights $w_s$ and $w_p$ in Equation~\ref{eq:reward} are set to 0.65 and 0.35, respectively. We emphasize more on the size reduction term here because bonuses in Table~\ref{tab:bonus} consider more on push quality. The qualitative magnitudes used in the bonus term are set to: $\gamma_H = 0.4$, $\gamma_M = 0.2$, and $\gamma_L = 0.1$. The exploration level $\alpha = 1.0$. The penalty factor in size reduction is $\beta=1.5$. The penalty in push quality $p_p=0.1$. All reward- and bandit-related hyperparameters are fixed globally across the entire benchmark suite to keep a consistent reward scale.

The benchmark suite comprises a total of 914 model checking problems from HWMCC'20, HWMCC'24, and HWMCC'25~\cite{HWMCC20Benchmarks,Preiner2025HWMCC25,Preiner2024HWMCC24}. 
To demonstrate the generality of the proposed method, experiments were conducted using  rIC3~\cite{ric3} (\href{https://github.com/gipsyh/rIC3/tree/8745148744448206ce8caef68beaea2bbecf57a3}{v1.3.6}). rIC3 is the champion of both bit-level track and word-level track: bit-vectors in HWMCC'24 and HWMCC'25~\cite{HWMCC24,Preiner2025HWMCC25}.  

The baselines in our experiments are the default rIC3 setting (denoted as rIC3-Standard), the rIC3 implementation of CtgDown~\cite{CTG} (denoted as rIC3-CtgDown), and the rIC3 DynAMic setting~\cite{EXCTG} (denoted as rIC3-DynAMic). The PA-LinUCB implementation utilizes the Rust crate \texttt{nalgebra} for linear algebra operations, ensuring efficient matrix computations across both platforms. The timeout limit for each model checking problem is 3600 seconds in wall-clock time. \looseness=-1

\subsection{MAB Configuration}
\label{sec:mab-config}


The configuration of the \AICTHREE is derived from extensions of the rIC3-DynAMic strategy.

\textbf{Arm Design.}
An arm instantiates a concrete setting of three CTG-related parameters:
$\mathsf{ctgMax}$, $\mathsf{ctgDepth}$, and
$\mathsf{exctgBudget}$.
They denote the maximum number of CTG attempts, the CTG recursion depth,
and the number of EXCTG block operations, respectively.
We use one basic static arm, three static arms with fixed settings, and three activity-aware \emph{dynamic} arms.
Our dynamic arms are extended from the activity-based parameterization in DynAMic~\cite{EXCTG}:
We keep the same piecewise structure, use the original mapping as the balanced dynamic arm, and add two variants.
(aggressive/conservative) by lowering/raising the activity thresholds and adjusting the scaling factors.
Table~\ref{tab:arms-both} summarizes all arms used in our experiments.

\begin{table}[h]
\centering
\caption{Arms configurations for \AICTHREE and baselines. 
}
\label{tab:arms-both}
\begin{tabular}{|c|l|l|ccc|}
\hline
Configuration &Arm & Type & $\mathsf{ctgMax}$ & $\mathsf{ctgDepth}$ & $\mathsf{exctgBudget}$ \\
\hline
\multirow{7}{*}{\AICTHREE (Ours)} 
& Basic       & static  & 0 & 0 & 0 \\
& Conservative & static  & 1 & 3 & 1 \\
& Balanced     & static  & 2 & 5 & 1 \\
& Aggressive   & static  & 8 & 4 & 1 \\
& Conservative & dynamic & \multicolumn{3}{c|}{$f_{\text{cons}}(a)$ (Equation~\ref{eq:dyn-cons})} \\
& Balanced     & dynamic & \multicolumn{3}{c|}{$f_{\text{bala}}(a)$ (Equation~\ref{eq:dyn-balanced})} \\
& Aggressive   & dynamic & \multicolumn{3}{c|}{$f_{\text{aggr}}(a)$ (Equation~\ref{eq:dyn-aggr})} \\
\hline
rIC3-Standard & \multicolumn{1}{c|}{\textemdash } & static & 0 & 0 & 0 \\ \hline
rIC3-CtgDown  & \multicolumn{1}{c|}{\textemdash } & static & 1 & 3 & 0 \\ \hline
rIC3-DynAMic  & \multicolumn{1}{c|}{\textemdash } & dynamic & \multicolumn{3}{c|}{$f_{\text{bala}}(a)$ (Equation~\ref{eq:dyn-balanced})} \\ \hline
\end{tabular}
\end{table}


\textbf{Dynamic arms.}
Following DynAMic~\cite{EXCTG}, an activity score $a$ is used to characterize the difficulty of blocking the current CTI (higher $a$ indicates that stronger generalization is needed).
Based on $a$, DynAMic~\cite{EXCTG} defines an activity-aware mapping to produce
$(\mathsf{ctgMax},\mathsf{ctgDepth},\mathsf{exctgBudget})$; we adopt this mapping as our balanced dynamic arm.
In addition, we introduce aggressive and conservative variants that keep the same three-regime
(low/medium/high activity) form, but use lower/higher thresholds and scaling to
encourage/discourage stronger generalization, respectively:
\begin{align}
\label{eq:dyn-balanced}
f_{\text{bala}}(a) &=
\begin{cases}
(0,0,0), & a < 10,\\
(\lfloor (a-10)/10 \rfloor + 2,\ 1,\ 1), & 10 \le a < 40,\\
(5,\ 1,\ \operatorname{round}((a-40)^{0.3}\cdot 2 + 5)), & a \ge 40,
\end{cases}\\[2pt]
\label{eq:dyn-aggr}
f_{\text{aggr}}(a) &=
\begin{cases}
(1,1,1), & a < 5,\\
(\lfloor (a-5)/8 \rfloor + 3,\ 1,\ 2), & 5 \le a < 25,\\
(6,\ 1,\ \operatorname{round}((a-25)^{0.3}\cdot 2.5 + 6)), & a \ge 25,
\end{cases}\\[2pt]
\label{eq:dyn-cons}
f_{\text{cons}}(a) &=
\begin{cases}
(0,0,0), & a < 15,\\
(\min(\lfloor (a-15)/12 \rfloor + 1, 3),\ 0,\ 1), & 15 \le a < 50,\\
(3,\ 1,\ \min(\operatorname{round}((a-50)^{0.3}\cdot 1.5 + 4), 6)), & a \ge 50.
\end{cases}
\end{align}

\definecolor{lightblue}{RGB}{173, 216, 230}
\subsection{Results}

The evaluation is organized around three questions. 
First, we compare \AICTHREE implementation on rIC3 with the baseline rIC3 configurations in terms of solved instances and runtime.
Second, we inspect the arm-selection behavior of \AICTHREE to understand whether the learned controller reduces to a single fixed configuration or adapts across different proof phases. 
Third, we analyze the termination frame level to examine whether the performance improvement is accompanied by faster proof convergence.

\subsubsection{Overall Performance}

\begin{table}[htbp]
\centering
\caption{Summary of results. 
}
\label{tab:ric3_bandit}

\begin{tabular}{|l|c|c|c|c|c|c|c|c|c|}
\hline
Configuration & \#Solved & Safe & Unsafe & $\Delta$ & $\Delta_s$ & $\Delta_u$ & PAR-2 & PAR-2\textsuperscript{\dag} & Average\textsuperscript{\dag} \\
\hline
rIC3-Standard     & 603 & 464 & 139 & 0   & 0   & 0   & 2541.05 & 5214.20 & 2953.17 \\
rIC3-CtgDown      & 629 & 478 & 151 & +26 & +14 & +12 & 2336.71 & 4794.40 & 2848.64 \\
rIC3-DynAMic      & 627 & 479 & 148 & +24 & +15 & +9  & 2346.48 & 4815.32 & 2783.76 \\
\AICTHREE (Ours) &
\cellcolor{lightblue}653 &
\cellcolor{lightblue}499 &
\cellcolor{lightblue}154 &
\cellcolor{lightblue}+50 &
\cellcolor{lightblue}+35 &
\cellcolor{lightblue}+15 &
\cellcolor{lightblue}2151.76 &
\cellcolor{lightblue}4413.68 &
\cellcolor{lightblue}2755.06 \\
\hline

\end{tabular}

{\footnotesize\textsuperscript{\dag} Calculated for non-trivial cases (cases unsolvable within 100 seconds).}

\end{table}

\begin{table}[t]
\centering
\small
\caption{Per-benchmark PAR-2 breakdown.}
\label{tab:per-benchmark-par2}
\setlength{\tabcolsep}{6pt}
\begin{tabular}{|l|c|c|c|}
\hline
Configuration & HWMCC'20 & HWMCC'24 & HWMCC'25 \\
\hline
rIC3-Standard & 1872.02 & 3210.70 & 2586.63 \\
rIC3-CtgDown  & 1719.38 & 3015.54 & 2309.31 \\
rIC3-DynAMic  & 1617.83 & 3061.78 & 2376.00 \\
\AICTHREE     & \cellcolor{lightblue}1479.82 & \cellcolor{lightblue}2821.75 & \cellcolor{lightblue}2200.99 \\
\hline
\end{tabular}
\end{table}

Table~\ref{tab:ric3_bandit} summarizes the experimental results on rIC3 model checker under four generalization configurations: rIC3-Standard, rIC3-CtgDown~\cite{CTG}, rIC3-DynAMic~\cite{EXCTG}, and our \AICTHREE. The table reports the number of solved instances (\#Solved), broken down into Safe and Unsafe, as well as the improvement over the baseline ($\Delta$, $\Delta_s$, $\Delta_u$) for the whole benchmark suite. We also report PAR-2, and for non-trivial cases (instances not solvable within 100 seconds), the corresponding PAR-2\textsuperscript{\dag} and average runtime\textsuperscript{\dag}.
  The PAR-2 score is a commonly used metric in competitions. It is calculated as the average time taken to solve the instances, with the timeout cases counted as twice the time limit. A lower PAR-2 score indicates better performance.  
  Overall, \AICTHREE achieves the best performance: it solves 653 instances in total (+50 over Standard, with +35 Safe and +15 Unsafe), and it also yields the lowest PAR-2 and PAR-2\textsuperscript{\dag} scores together with the lowest average runtime on non-trivial cases, outperforming both rIC3-CtgDown and rIC3-DynAMic across all reported metrics. 
  Table~\ref{tab:per-benchmark-par2} further reports the PAR-2 breakdown on HWMCC'20~\cite{HWMCC20Benchmarks}, HWMCC'24~\cite{Preiner2024HWMCC24}, and HWMCC'25~\cite{Preiner2025HWMCC25} after duplicate removal. \AICTHREE obtains the best PAR-2 on all three benchmark families.

Notably, \AICTHREE yields a more pronounced improvement on Safe instances than the other two IC3 variants, suggesting that our method can better incorporate the proving context to guide generalization and thus accelerate convergence on safety proofs. 
In particular, considering that rIC3 was the champion of the latest HWMCC~\cite{HWMCC24,Preiner2025HWMCC25}, the fact that our method can deliver further improvements is especially noteworthy.

Figure~\ref{fig:scatter_ric3_ic3ref} illustrates the performance comparison on the benchmark suite when running rIC3 with and without MAB. We observe a cluster of instances in the green region. These cases are effectively intractable for the baseline configurations within the time limit, yet become solvable once MAB is enabled. This indicates that adaptively selecting a context-appropriate generalization strategy can unlock substantial speedups, sometimes by orders of magnitude, on originally hard instances. Meanwhile, on moderately hard cases where multiple strategies are already viable, \AICTHREE closely tracks the baseline runtime, suggesting that the additional exploration introduced by MAB incurs little overhead in practice.

\newcommand{\sfw}{0.31\textwidth} 

\begin{figure}[htbp]
  \centering

  \begin{minipage}[b]{\sfw}\centering
    \includegraphics[width=\linewidth,
    alt={Scatter plot comparing CPU time of \AICTHREE against rIC3-CtgDown on a logarithmic scale. The x-axis is rIC3-CtgDown CPU time in seconds and the y-axis is \AICTHREE CPU time in seconds. Points are marked as Safe/Proof, Unsafe/CEX, or Unknown. A diagonal reference line indicates equal runtime, and the shaded region highlights hard instances where rIC3-CtgDown is slow while \AICTHREE is faster.}]{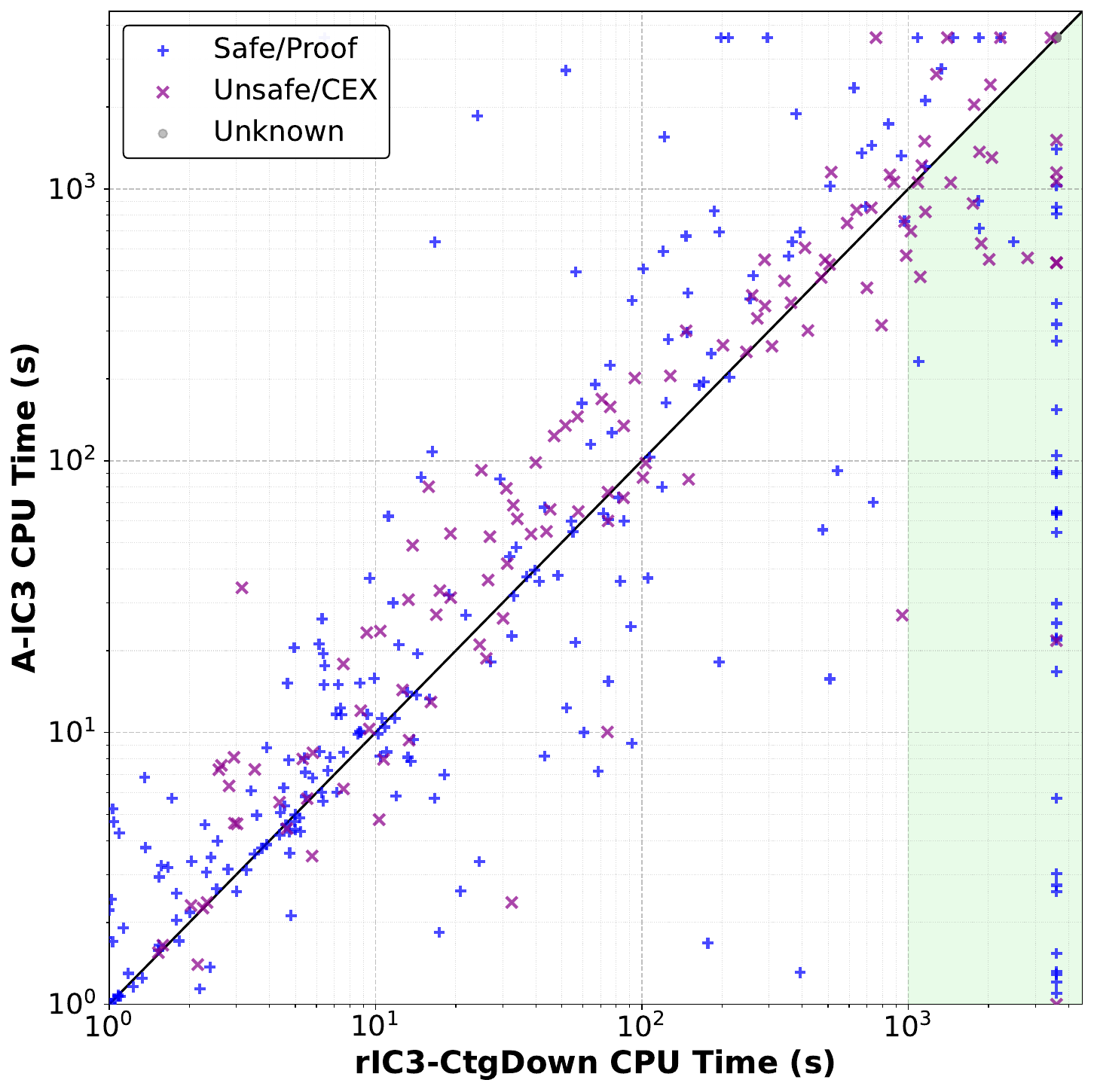}
    \par\small (a)
  \end{minipage}\hfill
  \begin{minipage}[b]{\sfw}\centering
    \includegraphics[width=\linewidth,
    alt={Scatter plot comparing CPU time of \AICTHREE against rIC3-DynAMic on a logarithmic scale. The x-axis is rIC3-DynAMic CPU time in seconds and the y-axis is \AICTHREE CPU time in seconds. Points are marked as Safe/Proof, Unsafe/CEX, or Unknown. A diagonal reference line indicates equal runtime, and the shaded region highlights hard instances where rIC3-DynAMic is slow while \AICTHREE is faster. Points in shaded region are fewer than that of rIC3-CtgDown's}]{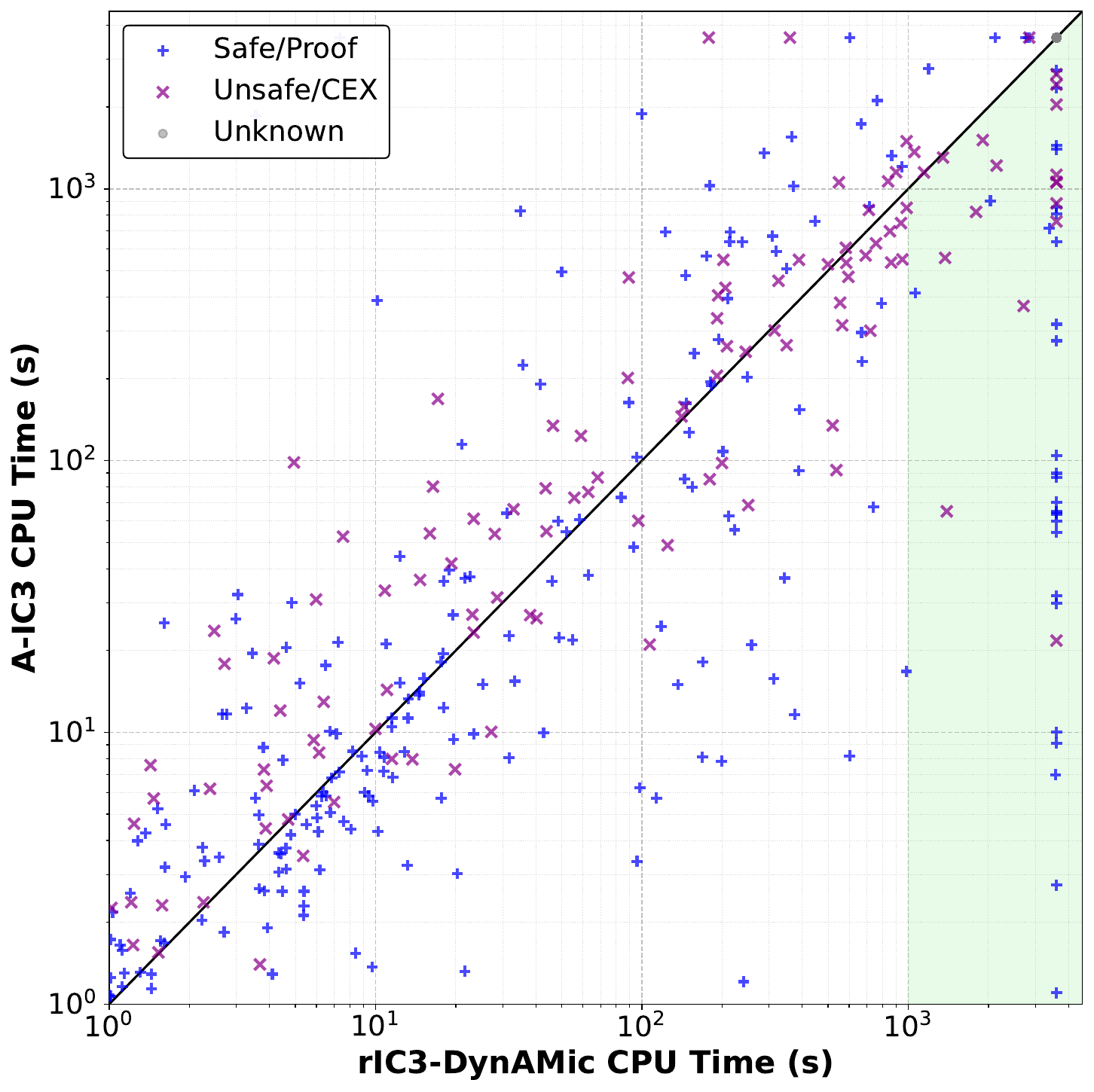}
    \par\small (b)
  \end{minipage}\hfill
  \begin{minipage}[b]{\sfw}\centering
    \includegraphics[width=\linewidth,
    alt={Scatter plot comparing CPU time of \AICTHREE against rIC3-Standard on a logarithmic scale. The x-axis is rIC3-Standard CPU time in seconds and the y-axis is \AICTHREE CPU time in seconds. Points are marked as Safe/Proof, Unsafe/CEX, or Unknown. A diagonal reference line indicates equal runtime, and the shaded region highlights hard instances where rIC3-Standard is slow while \AICTHREE is faster. Points in shaded region are more than that of rIC3-CtgDown's}]{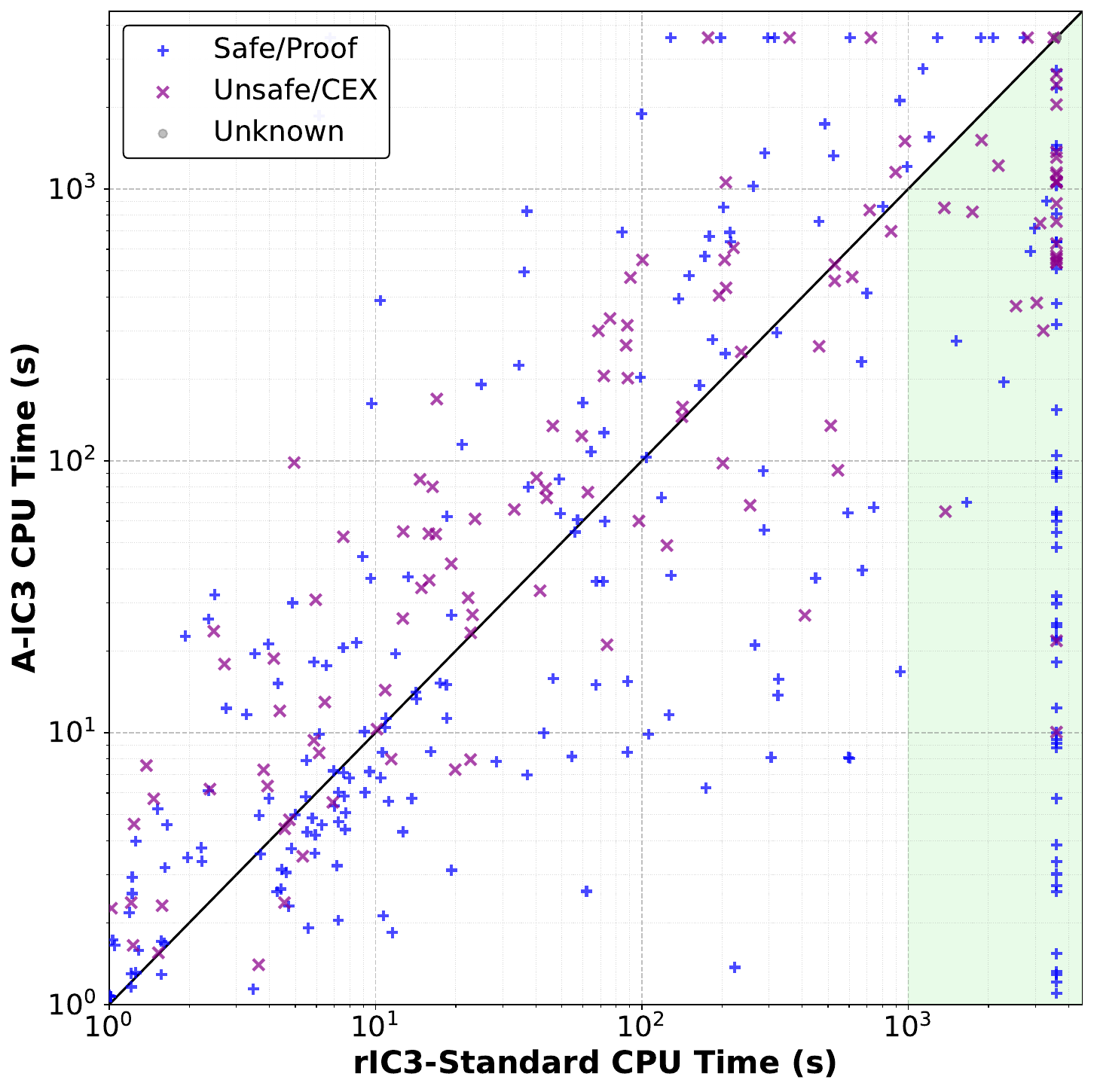}
    \par\small (c)
  \end{minipage}

  \vspace{1mm}


  \caption{Performance comparison of rIC3 with and without MAB: \AICTHREE (with MAB) vs.\ non-MAB baselines.
(a) \AICTHREE vs. rIC3-CtgDown;
(b) \AICTHREE vs. rIC3-DynAMic;
(c) \AICTHREE vs. rIC3-Standard.
Green regions indicate hard instances where the baseline struggled, but \AICTHREE achieved better results.}
  \label{fig:scatter_ric3_ic3ref}
\end{figure}

\begin{figure}[htbp]
  \centering
  \begin{minipage}[b]{0.49\textwidth}
    \centering
    \includegraphics[width=\linewidth,
    alt={Cactus plot comparing the number of solved benchmark cases over time for rIC3-Standard, rIC3-CtgDown, rIC3-DynAMic, and \AICTHREE. The x-axis is the time threshold in seconds from 0 to 3600, and the y-axis is the cumulative number of solved cases. The MAB-based configuration stays above the other curves across most time thresholds and reaches the highest final solved count.}]{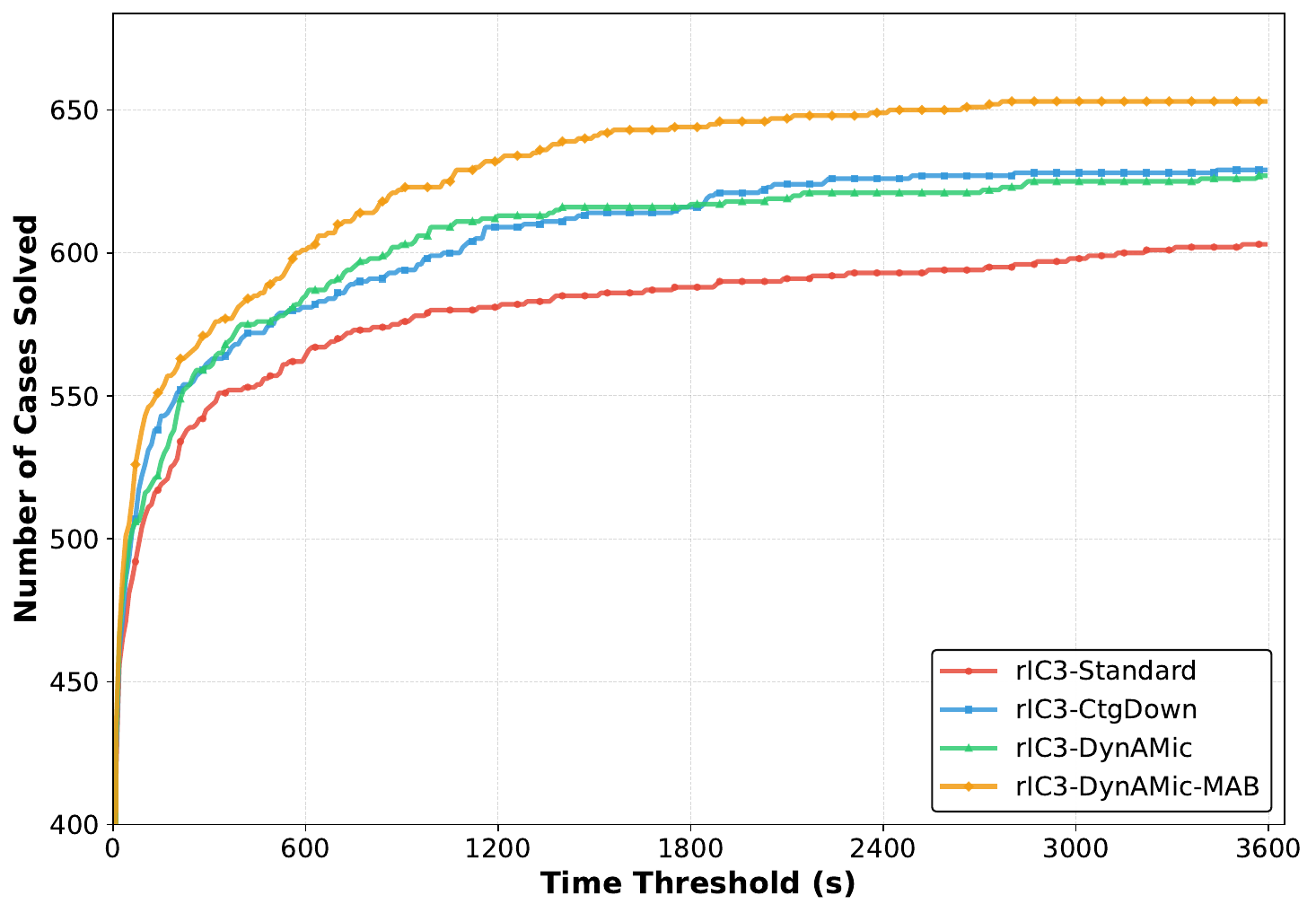}
  \end{minipage}\hfill

  \caption{Cactus plots comparing solver performance. 
  }
  \label{fig:cactus_ric3_ic3ref}
\end{figure}

 Figure~\ref{fig:cactus_ric3_ic3ref} shows the number of cases solved over time. The MAB-enhanced method consistently dominates the other configurations across the entire time range. Beyond the final number of solved instances, the cactus curves also highlight how MAB improves the efficiency of proof construction. The MAB curve rises faster in the early and middle time budgets, indicating that it reduces time spent on unpromising generalization choices and reaches effective proof trajectories sooner. As the timeout threshold increases, the gap remains and even expands for harder instances, suggesting that adaptive strategy selection continues to pay off when the proof search becomes more sensitive to generalization quality. At the same time, the curve does not exhibit a noticeable slowdown on easier and moderately hard cases, which supports our claim that the exploration overhead is well controlled and does not negate the gains from selecting better strategies.

\subsubsection{Arm Selection Behavior}

\begin{figure}[htbp]
    \centering
    \includegraphics[width=0.98\linewidth,
    alt={Stacked bar charts showing the percentage of \AICTHREE arm selections over normalized solving progress for three benchmark instances: xepic\_a17-p02, yosyshq_apnote\_123\_cv32e40x-p142, and zipcpu-zipmmu-p00. Each chart divides solving progress into five intervals from 0--20\% to 80--100\%, and each bar is split among Basic, Conservative Static, Conservative Dynamic, Balanced Static, Balanced Dynamic, Aggressive Static, and Aggressive Dynamic arms. The distributions change over time and differ across instances, illustrating adaptive strategy selection during solving.}]{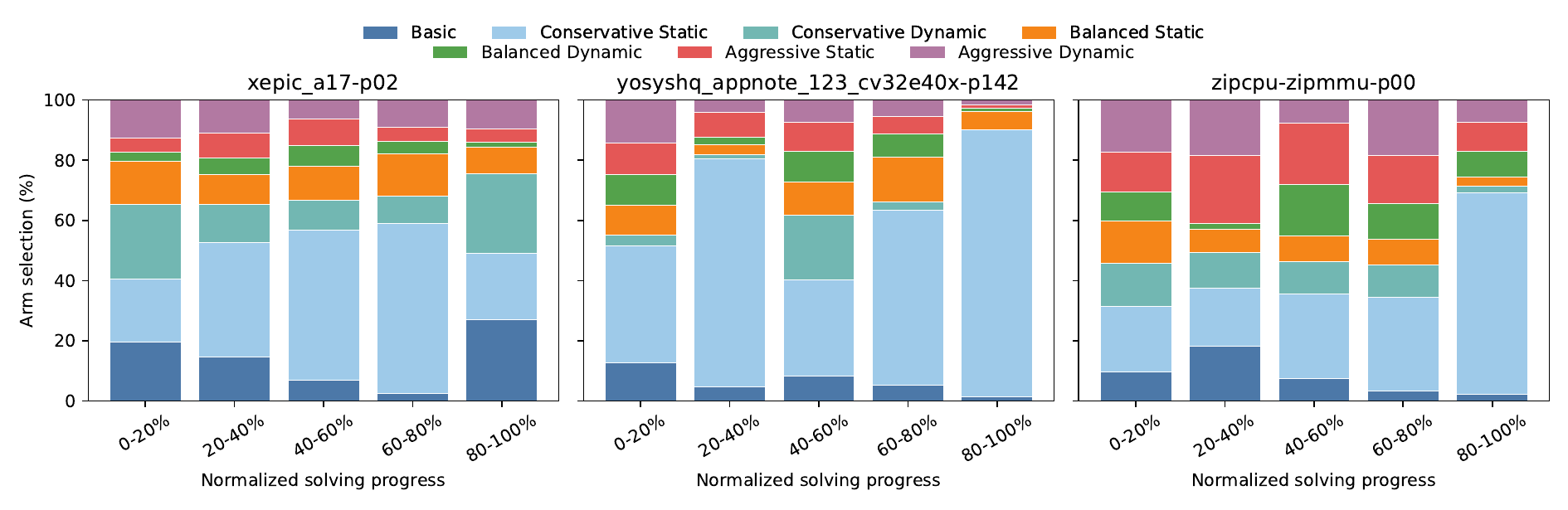}
    \caption{Arm selection distribution over proof phases on case studies.}
    \label{fig:arm_selection}
\end{figure}

To further examine whether the improvement comes from adaptive strategy selection rather than from repeatedly favoring one fixed configuration, we
analyze the arm choices made by \AICTHREE during the solving process.
Figure~\ref{fig:arm_selection} reports the arm-selection distribution on three representative non-trivial instances where \AICTHREE obtains clear improvements over the baseline configurations. 
For each run, we divide the sequence of generalization calls into five equal phases according to normalized solving progress, and report the percentage of calls in which each arm is selected.
These cases are used only for mechanism-level illustration; the aggregate
performance claims are based on the full benchmark suite.

The results show that the dominant arms vary across both instances and solving phases. For example, on \texttt{xepic\_a17-p02}, the selection gradually shifts from a mixed use of conservative and balanced arms toward a larger fraction of basic and conservative arms in the last phase. 
On \texttt{yosyshq\_appnote\_123\_cv32e40x-p142}, conservative static generalization dominates most phases, but the early and middle phases still involve non-negligible selections of other static and dynamic arms. 
On \texttt{zipcpu-zipmmu-p00}, the first four phases use a more diverse mixture of arms, whereas the last phase is dominated by conservative  generalization.
Overall, these observations suggest that \AICTHREE re-ranks the arm space throughout the run according to the current proof context and accumulated reward feedback, rather than following a manually predefined switching schedule with better performance.
In these successful non-trivial cases, the learned controller often allocates more effort to broader or deeper CTG exploration in the early and middle phases, and then shifts toward shallower choices near the end of solving to reduce frame saturation and over-generalization.

\newcommand{\lvlfigdir}{fig/cav_fig_24}

\subsubsection{Effect on Proof Convergence}

\begin{figure}[htbp]
  \centering

  \begin{minipage}[b]{\sfw}\centering
    \includegraphics[width=\linewidth,
    alt={Comparison of IC3 termination level count between \AICTHREE and rIC3-CtgDown on a logarithmic scale. The upper scatter plot uses rIC3-CtgDown level count on the x-axis and \AICTHREE level count on the y-axis, with a dashed diagonal line indicating equal level count. Points are marked as better, worse, or equal performance for \AICTHREE. The lower plot shows the distribution of the level-count ratio \AICTHREE divided by rIC3-CtgDown, separating cases where \AICTHREE performs better or worse. Most better-performance points have ratios below one, indicating that \AICTHREE often terminates with fewer levels in case of better performance.}]{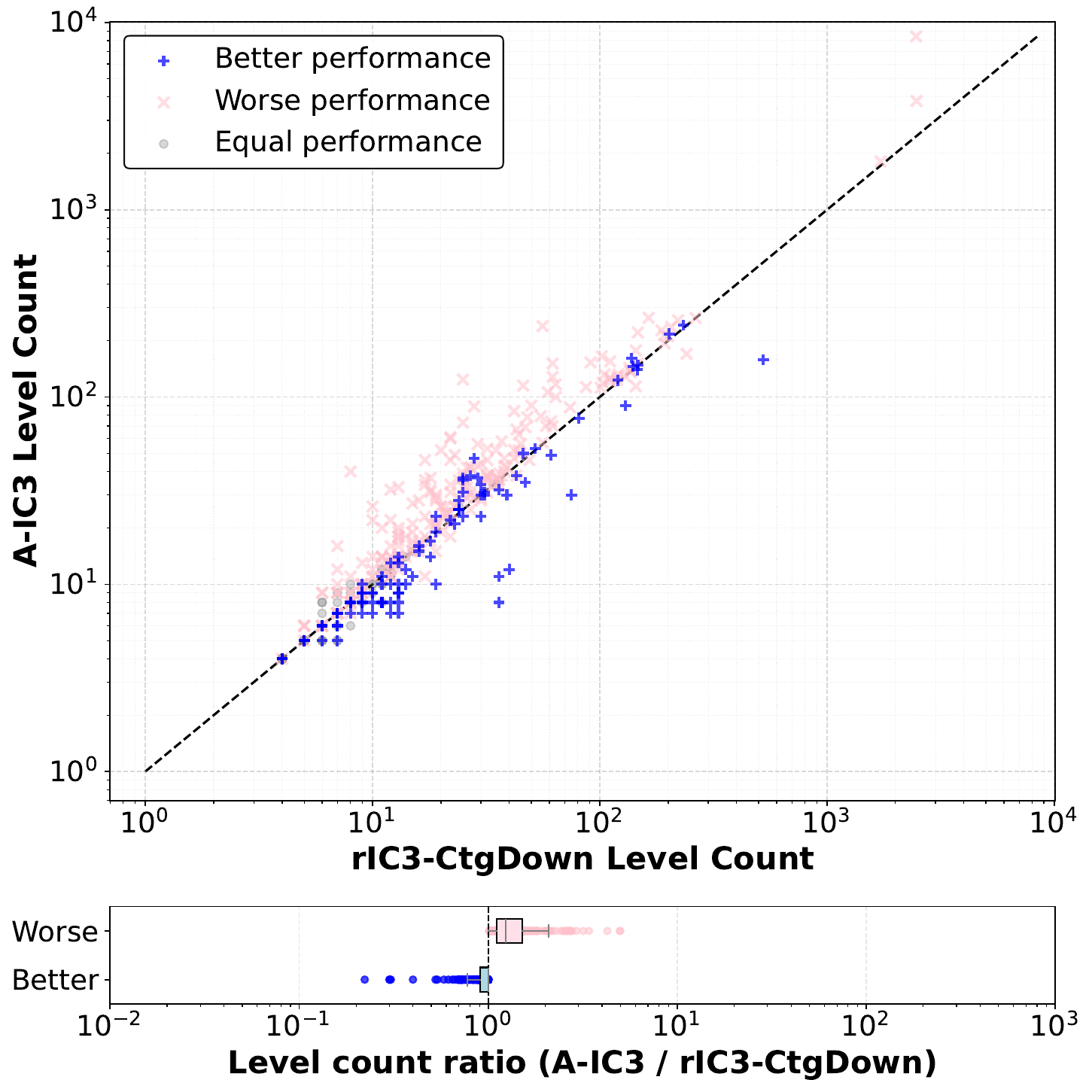}
    \par\small (a)
  \end{minipage}\hfill
  \begin{minipage}[b]{\sfw}\centering
    \includegraphics[width=\linewidth,
    alt={Comparison of IC3 termination level count between \AICTHREE and rIC3-DynAMic on a logarithmic scale. The upper scatter plot uses rIC3-DynAMic level count on the x-axis and \AICTHREE level count on the y-axis, with a dashed diagonal line indicating equal level count. Points are marked as better, worse, or equal performance for \AICTHREE. The lower plot shows the distribution of the level-count ratio \AICTHREE divided by rIC3-DynAMic, separating cases where \AICTHREE performs better or worse. Most better-performance points have ratios below one, indicating that \AICTHREE often terminates with fewer levels than rIC3-DynAMic in case of better performance.}]{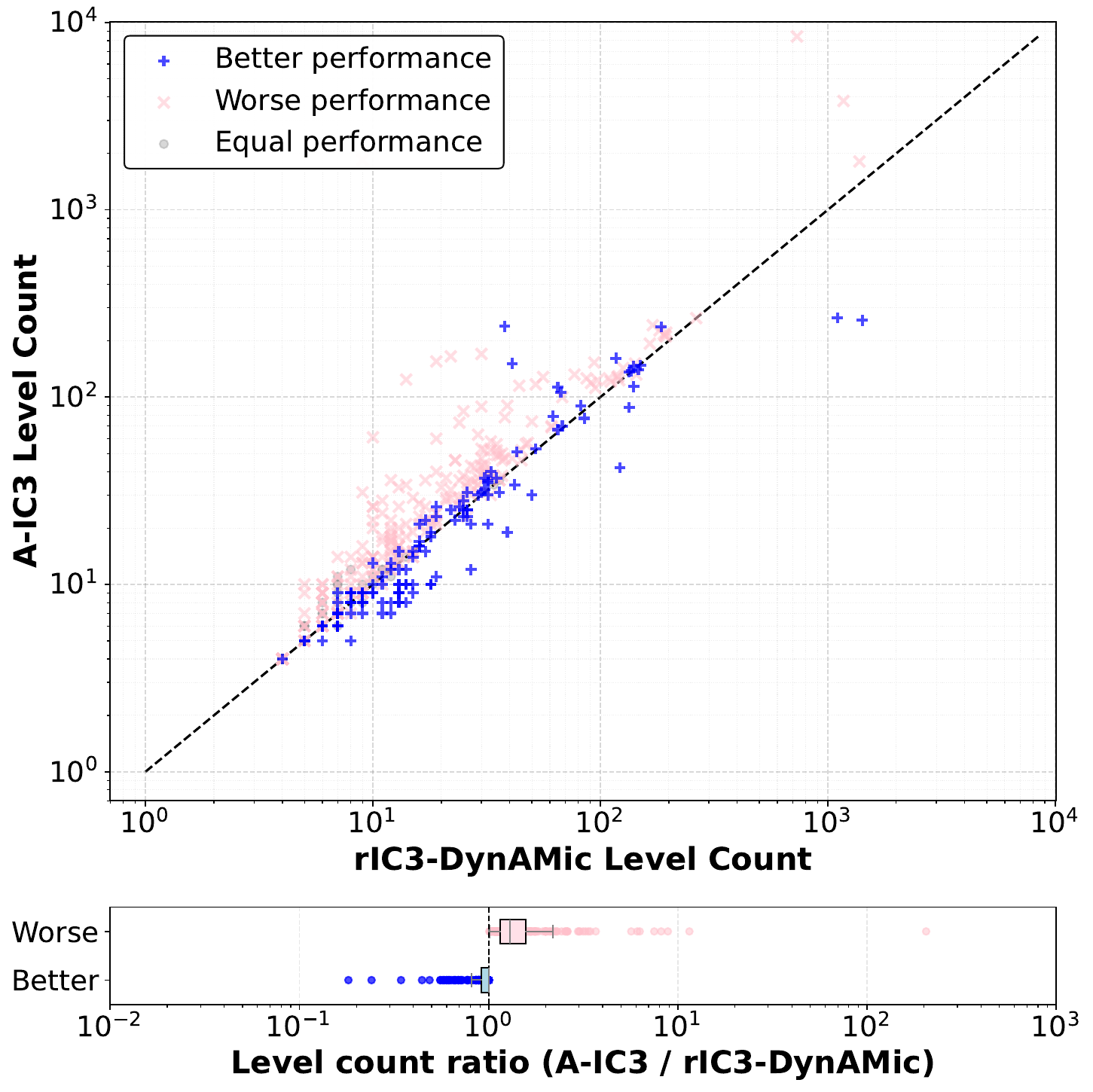}
    \par\small (b)
  \end{minipage}\hfill
  \begin{minipage}[b]{\sfw}\centering
    \includegraphics[width=\linewidth,
    alt={Comparison of IC3 termination level count between \AICTHREE and rIC3-Standard on a logarithmic scale. The upper scatter plot uses rIC3-Standard level count on the x-axis and \AICTHREE level count on the y-axis, with a dashed diagonal line indicating equal level count. Points are marked as better, worse, or equal performance for \AICTHREE. The lower plot shows the distribution of the level-count ratio \AICTHREE divided by rIC3-Standard, separating cases where \AICTHREE performs better or worse. Most better-performance points have ratios below one, indicating that \AICTHREE often terminates with fewer levels than rIC3-Standard in case of better performance.}]{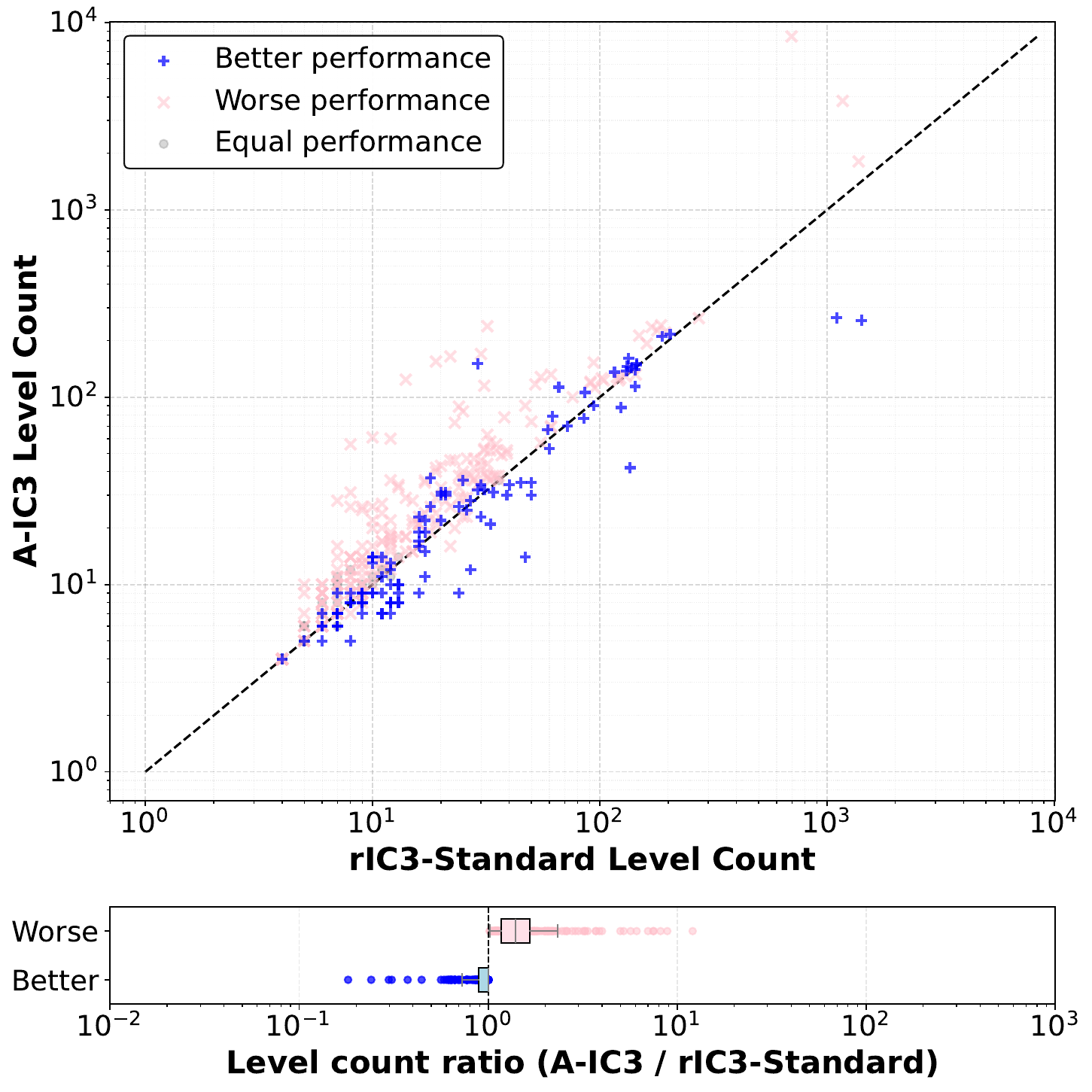}
    \par\small (c)
  \end{minipage}

  \vspace{1mm}


  \caption{Comparison of the termination level count. 
  (a) \AICTHREE vs.\ rIC3-CtgDown;
  (b) \AICTHREE vs.\ rIC3-DynAMic;
  (c) \AICTHREE vs.\ rIC3-Standard.
  }
  \label{fig:convergence_level}
\end{figure}

Figure~\ref{fig:convergence_level} compares the termination level count (the index/number of frames when the solver terminates, either by reaching an inductive invariant or by finding a counterexample) of \AICTHREE against the baselines. Marker \texttt{$\times$} denotes instances where the MAB-enhanced solver performs worse than the baseline, while \texttt{+} denotes instances where it performs better. It can be observed that a higher termination level generally indicates poorer performance. Across all three comparisons in Figure~\ref{fig:convergence_level}, the overwhelming majority of points lie on or below the diagonal, and the level-count ratios are mostly below 1, indicating that MAB typically terminates at a lower level than the baselines, rather than improving only a small set of outliers. Since a lower convergence level in IC3 usually means fewer frames are needed to make the proof inductive or to expose a counterexample, this trend suggests that MAB accelerates solving by improving the proof trajectory itself. This observation also echoes our earlier claim that no single generalization heuristic is uniformly optimal: by adapting the strategy to the current proving context, MAB tends to derive clauses that are more ``pushable'' and better aligned with the proof frontier, which helps close the inductive gap earlier. Cases where MAB increases the termination level do exist, but they are relatively sparse and often close to the diagonal.

\section{Related Work}\label{sec:related_works}


\textbf{Heuristic-Based Generalization.}
A key line of research focuses on heuristics for improving clause quality. The seminal CTG~\cite{CTG} method enhances generalization by blocking states that impede the process (CTGs). Its extension, EXCTG~\cite{EXCTG}, further improves generalization by recursively blocking predecessors of CTGs. While effective, these methods rely on fixed or reactive heuristics (e.g., DynAMic in~\cite{EXCTG}), which may not adapt to the global proof state. \AICTHREE instead uses a richer context to anticipate the optimal strategy.

\textbf{Global Guidance in Model Checking.}
Another direction complements local generalization with global guidance. GSPACER~\cite{vediramana2024global} uses theory-agnostic rules (Subsume, Concretize, Conjecture) to steer the search based on the global set of lemmas, mitigating issues like myopic generalization. GSPACER-BV~\cite{vk2020word} extends this to bit-vectors, using global guidance as a substitute for unavailable interpolation methods. These works focus on \textit{what} to generalize. In contrast, \AICTHREE is orthogonal and focuses on \textit{how} to generalize by adaptively selecting the best procedure.

\textbf{Data-Driven and Learning-Based Approaches.}
Machine learning has also been applied to guide generalization. ROPEY~\cite{le2021data} uses a recurrent network to predict which literals to keep, while NeuroPDR~\cite{NeuroPDR} and DeepIC3~\cite{DEEPIC3} employ Graph Neural Networks (GNNs) to predict effective clauses from the circuit structure. Although powerful, these supervised methods require extensive offline training and can introduce significant runtime overhead. \AICTHREE sidesteps these issues with an MAB formulation that learns online with minimal overhead and no pre-existing data. It acts as a lightweight, adaptive control layer, making it more dynamic than fixed heuristics and more practical for hardware model checking than heavyweight deep learning approaches.

\textbf{RL-Guided SAT/SMT Solving.}
Bandit and RL techniques have been explored for guiding SAT/SMT solving, including MAB-style branching heuristics such as CHB~\cite{Liang2016CHB} and LRB~\cite{Liang2016LRB}, MAB-based switching of SAT branching heuristics~\cite{Cherif2021VSIDSCHB}, and MAB-guided SMT solver~\cite{Pimpalkhare2021MedleySolver} or quantifier-instantiation selection~\cite{Jakubuv2023QuantifierBandits}. 
Since IC3 relies on SAT queries for reachability and relative-inductiveness checks, these techniques could improve the underlying SAT calls. 
In contrast, \AICTHREE operates at the IC3 algorithm layer rather than inside the SAT/SMT solver and is complementary to bandit-guided SAT/SMT solving.

\section{Conclusion}\label{sec:conclusion}

In this work, we presented a lightweight, learning-guided framework for adaptive generalization in IC3. By modeling strategy selection as a contextual multi-armed bandit problem, our method dynamically adjusts generalization to the proof context, overcoming static heuristic limitations. Experiments demonstrated improvements over baseline strategies on extensive benchmarks. These results highlight the potential of integrating machine learning into formal verification for more efficient model checking. Future work includes deeper integration with other verification phases and advanced reinforcement learning methods.




\begin{credits}

\subsubsection{\ackname}
This work is supported by the Guangdong S\&T Program (No. 2025A0505000022)
and the Hong Kong Research Grants Council General Research Fund (GRF)
(No. 16218324). 
We thank Yuheng Su for his helpful discussions and support for this work.
Hongce Zhang and Wei Zhang are co-corresponding authors.

\subsubsection{\discintname}
The authors have no competing interests to declare that are relevant to the
content of this article.

\end{credits}

\subsubsection*{Data-Availability Statement}
Our implementation is available at:
\url{https://zenodo.org/records/19811923}.
%
%
%
%
\bibliographystyle{splncs04}
\bibliography{MAB-IC3.bib}
\end{document}